\documentstyle[twocolumn,prb,aps,epsf]{revtex}

\begin{document}
\draft

\wideabs{
\title{\bf NONADIABATIC EFFECTS IN A GENERALIZED JAHN-TELLER LATTICE
MODEL: HEAVY AND LIGHT POLARONS, PAIRING AND METAL-INSULATOR TRANSITION}
\author{ Eva Majern\'{\i}kov\'a${}^{\dag \ddag *}$, J.
Riedel${}^{\dag}$
and S. Shpyrko${}^{\dag **}$ }
\address{${}^{\dag}$Department of Theoretical Physics, Palack\'y University, \\
T\v r. 17. listopadu 50, CZ-77207 Olomouc, Czech Republic \\
${}^{\ddag}$Institute of Physics, Slovak Academy of Sciences, \\
D\'ubravsk\'a cesta, SK-84 228 Bratislava, Slovak Republic
}

\maketitle

%%%%%%%%%%%%%%%%%%%%%%%%%%%%%%%%%%%%%%%%%%%%%%%%%%%%%%%%%%%%%%%%%%
\begin{abstract}
The selfconsistent ground state polaron potential
of one-dimensional lattice of  two-level molecules with spinless
electrons and two  dispersionless phonon modes with linear coupling and
quantum phonon-assisted (nonadiabatic)
transitions between the levels is found anharmonic in phonon
displacements. As a function of these, the potential shows a crossover from
two nonequivalent broad minima to a single narrow minimum which correspond
to the positions of the levels in the ground state.
Generalized variational approach respecting the mixing of levels
(reflection) via a variational parameter implies prominent nonadiabatic effects:
 (i) In the limit of the symmetric E$\otimes$e Jahn-Teller situation
 they cause transition between the regime of the predominantly one-level
 "heavy" polaron and a "light" polaron oscillating between the levels
 due to phonon assistance with almost vanishing polaron displacement.
  Vanishing polaron selflocalization implies
  {\it enhancement of the electron transfer} due to
  decrease of the "heavy" polaron mass (undressing) at the point of
  the transition.
 There can occur {\it pairing of "light" polarons}
 due to exchange of virtual phonons. {\it Continuous
 transition to new energy ground state close to the transition
 from "heavy" polaron phase to "light" (bi)polaron phase occurs}.
 In the "heavy" phase, we have found anomalous (anharmonic)
enhancements of quantum fluctuations of the
phonon coordinate, conjugated momentum and their product in the ground state
as functions of the effective coupling which reach their maxima at
E$\otimes $e JT symmetry. They decrease rapidly to their harmonic values as soon as the
"light" phase is stabilized.
(ii)  Nonadiabatic dependence of the polaron mass (Debye-Waller screening)
on the optical phonon frequency appears.
 (iii) The contribution of Rabi oscillations to the transfer enhances
 significantly quantum shift of the insulator-metal transition
line to higher values of the critical effective electron-phonon coupling
 supporting so the metallic phase. In the
E$\otimes$e JT case, insulator-metal transition can coincide
 with the transition between the "heavy" and the "light" (bi)polaron
 phase only at certain (strong) effective electron-phonon interaction.

\end{abstract}
\pacs{PACS number(s): 71.38.-k, 42.50.L, 31.30.Gs, 74.20.Mn}
}

\narrowtext
\section{Introduction}

Recently an interest in electron-phonon models was renewed owing to
high-$T_c$-superconducting layered cuprates and fullerene compounds
which exhibit structure instability due to strong Jahn-Teller effect
accompanied by an evident isotope effect
\cite{Kaplan:1995}$^{-}$\cite{Kumar:1998}.
Here, nonadiabatic (quantum) fluctuations were found to be important
since the majority of the superconducting structures exhibited low values
of the Fermi energy,
comparable to the phonon energy and large isotope effect.
Nonadiabatic fluctuations also appeared relevant
in manganese-based perovskites at Jahn-Teller distortion  assisted
by formation of JT polaron and causing oxygen isotope effect and a
collosal magnetoresistance \cite{Berlin:1997}.
 The mechanism of nonadiabatic pairing was proposed by several
 authors:  Manini et al \cite{Manini:1995}, Zheng et al
 \cite{Zheng:1989}$^{-}$\cite{Feinberg:1990}, Kresin et al
\cite{Bill:1998}$^{,}$\cite{Kresin:1994}, Pietronero et
al \cite{Pietronero:1992} for $C_{60}$ compounds  and recently
also for the challenging new superconductor $MgB_2$ \cite{Pietronero:2001}.
Theories of pairing mechanism based on polarons
(bipolarons), including also electron-electron interactions evoked 
great interest as well \cite{Alexandrov:1986}$^{-}$\cite{Wagner:1999}.

Nonadiabatic fluctuations affect the charge transport because they reduce the
polaron band narrowing \cite{Zheng:1989}$^{-}$\cite{Feinberg:1990}
(i.e. polaron renormalization of the electron mass).
Fluctuations increase near the phase transitions
 \cite{Fradkin:1983}$^{-}$\cite{Zheng:1996}
 destroying the phase coherence for weak interactions and shifting
the  critical couplings to higher values. However,
numerical simulations of different lattice models
 (Fradkin and Hirsch \cite{Fradkin:1983}, Borghi et al.
 \cite{Borghi:1995},  McKenzie et al \cite{McKenzie:1996})
 prove a general statement that the resulting  quantum fluctuations are much
 more  pronounced than those obtained by variational approaches,
 e.g. by Zheng et al. \cite{Zheng:1989}, Feinberg et al. \cite{Feinberg:1990},
 Lo\cite{Lo:1991} and Chen et al.\cite{Chen:1989}.
Namely, the quantum shift of  critical values
of the electron-phonon coupling from numerical simulations to higher values
(preferring so the metallic phase) was shown to exceed considerably
their variational values. Moreover, the numerical simulations of
 two-level lattice electron-phonon systems in one dimension
by Feinberg et al. \cite{Feinberg:1990} and by Borghi et al. \cite{Borghi:1995}
evidence anomalous increase of quantum fluctuations of the phonon
coordinate and of the conjugated momentum as well as anomalous
increase of their product referring to enhancement of anharmonic effects.
The simulations manifest a dramatic
increase of phonon fluctuations and of their product far beyond the
standard uncertainty principle.
 The inadequacy of the variational approaches was ascribed to
insufficiency of the squeezed coherent phonons (harmonic
oscillators) to comprise anharmonic behavior of the phonons in
two-level (-band) models.

Since the early beginning \cite{Jahn:1937} Jahn-Teller model was
investigated mainly in chemistry as a prototype
 model of electron-phonon interaction for localized centers in solids.
It was used to study the instability of the orbitally degenerate
electronic states of highly symmetric ionic
configuration against ionic distortions in
localized molecular centers in crystals \cite{Wagner:1984}$^{,}$
\cite{Obrien:1993}.
The effect was explained by Jahn-Teller theorem
about lifting the degeneracy of an orbitally degenerate electronic state
by symmetry lowering distortions of nuclear configurations.
This effect is also involved e.g. in
physics of structural phase transitions in solids doped with Jahn-Teller
active ions \cite{Gehring:1975}, for mechanisms based on
tunneling between two electronic levels coupled to phonon modes
\cite{Wagner:1984}, in optical
\cite{Obrien:1981} and paramagnetic ion spectra \cite{Ham:1972}.

Though JT effect was considered as the most representative
nonadiabatic system, Born-Oppenheimer approximation has been extensively used
although it is valid only in the limit of large local distortions.
Inconsistency of the adiabatic approach for small distortions
was analyzed by Wagner \cite{Wagner:1984}$^{,}$\cite{Wagner:1988}.

Importance of JT effect in physics was increased due to above mentioned
 discovery of JT effect based structural transitions in some
high-$T_c$ superconductors and in manganese-based perovskites.
Therefore, the consideration is now focused on lattice versions of the
JT model.

In this paper we will investigate an extended lattice of JT molecules:
site molecules with doubly degenerate electron level coupled to two internal
 phonon modes with different symmetries and different coupling constants.
The antisymmetric mode splits the levels, the symmetric one
 couples the levels via phonon assisted on-site and intersite electron
transitions (Rabi oscillations).
As we show, these transitions are especially important if the
difference of the effective potential minima as a function of the phonon
displacement is of the order of the phonon
energy (symmetric Jahn-Teller molecules).
We shall use a generalized variational approach
which will account {\it more aspects of nonadiabaticity}:
Instead of simple squeezed coherent phonons,
 as a more appropriate variational ansatz for phonons we take
 two-center squeezed coherent wave function
which accounts for possibility of the phonon mediated coupling of
levels.
The variational wave function of the two-level electron-phonon
systems  with reflection symmetry (antisymmetry) of the coherent phonon
states related to both the levels $\phi_+$ and  $\phi_-$ was first
introduced by Shore and Sander \cite{Sander:1973} as linear combinations
$\phi_+ + \eta \phi_-$ and $\phi_- + \eta \phi_+$, $\eta$ being
new (reflection) variational parameter.
The approach has been widely exploited and further developed by Wagner et al.
\cite{Wagner:1987}$^{,}$ \cite{Sonnek:1994}
 for exciton  ($(211)$), dimer ($(221)$ and $(222)$) systems
 (with the tunneling  between the levels in contrast to the phonon-assisted
 electron transitions of our model);  here the notation $(xyz)$ is used
 for $x$ as number of the levels, $y$ as the number of the sites,
 and  $z$ as number of the electrons in a cell).
 The detailed comparison of the ground state energies (GSE) of
 different  variational phonon wave functions for an exciton-phonon
 or dimer-phonon models \cite{Sonnek:1994}
 confirmed the two-peak variational choice
  as  the most suitable one, i.e. giving {\it the best fitting of the exact
  solution} in the medium and strong coupling regime.

Similar structure of the wave function was proposed for
the phonon wave function coupled to two electronic states of a
double-well potential by Kresin et al \cite{Kresin:1994} in
their model of the nonadiabatic origin of the isotope effect in high-$T_c$
superconductors.

  %%%%%%%%%%%%%%%%%%%%%%%%%%%%%%%%%%%%%%%%%%%%%%%%%%%%%%%%%%%%%%%%

In the Section II. we formulate variational approach for our
model. We find effective ground state potential as a function of
all (four) phonon variational
parameters (displacements of two phonon modes, squeezing and reflection
(mixing)).

In the Section III. we compare variational results for the ground
state energy of the
two-center squeezed coherent phonon wave function with the adiabatic
ground state energy and identify the region of importance of the
reflection ($\eta$) and nonsymmetry ($\chi=\beta/\alpha$) parameters to be the
 region close to $\chi=1$ (E$\otimes$e JT symmetry).
Special importance of the symmetric (Jahn-Teller) lattice molecules
occurs as a specific condition for which the reflection parameter is the
most effective.
At this symmetry, there takes place a nonadiabatic transition between
the polaron dominantly selflocalized within one
level,  "heavy" polaron of a broad minimum and the almost delocalized,
  "light" polaron of a narrow minimum (with vanishing displacement)
oscillating between close levels via both onsite and intersite phonon
assisted tunneling: Namely, the coherent phonons 2 are accompanied by
Rabi oscillations of the electron between the levels due to the phonon mode 1.
The latter virtual phonons mediate the coupling of polarons which may
occupy the levels.

We compare ground state characteristics of the "heavy" and "light"
region as functions of pairs of competing classical (effective coupling)
and quantum (phonon frequency, tunnelling) parameters.
 The self-trapping due to electron-phonon coupling
 competes with the lattice transfer of
electrons (band width) which itself is being renormalized by the
Debye-Waller factor.
The characteristics of the related insulator-metal transition, the value of
the gap and consequently the critical coupling are
determined by the complex interplay of the transfer supporting the metallic
phase and of the electron-phonon coupling supporting the
insulating phase.
The shift of the critical line to higher values of the critical
couplings due to quantum effects is discussed in the Section IV.

In the Section V. we investigate anomalous behavior of the squeezed
ground state quantum fluctuations of the canonically conjugated phonon
coordinates and their product on the effective coupling ($\mu$) and
effective potential asymmetry $(\chi)$ parameters,
namely the strong anharmonic fluctuations
which reach their maximum values again at $\chi=1$. They
decrease to the harmonic oscillator values  for $\chi > 1$.
A related model is a lattice of two-level
dimers with one spin electron at each site ($(222)$ lattice
model)  studied by exact numerical methods by Borghi et
al. \cite{Borghi:1995}. The present model differs from that one, but
concerning the quantum fluctuations of the phonon coordinate and momenta
qualitatively the same results  as for their
model are expected.

We remark that variational methods are widely used for
electron-phonon lattice models.
 
Fully analytic nonvariational approach with nonconservation
of the number of phonons was performed for a local ($D=0$) dimer by
Weber-Milbrodt \cite{Weber:1988}. Unfortunately, in the extended (D=1) lattice
model this method acquires extreme mathematical complexness. It was
used as a basis for numerical study \cite{Borghi:1995} of the
above mentioned lattice model.

\section{Extended (lattice) generalized Jahn-Teller model }

We investigate 1D lattice of spinless double degenerated electron
states linearly coupled to two intramolecular phonon modes
described by Hamiltonian

 \begin{eqnarray}
H= \Omega \sum_{n,i=1,2} (b_{in}^{\dag}b_{in} +1/2
 ) +\sum\limits_{n}
\left(  \alpha  (b_{1n}^{\dag}+b_{1,n})\sigma_{zn}
\right.\nonumber\\
\left. -\beta (b_{2n}^{\dag}+b_{2,n})\sigma_{xn}\right )
-\frac{T}{2} \sum_{n,j=1,2} (R_{1,j}+R_{-1,j})I_{n} .
\label{Hams1}
\end{eqnarray}

where $b_{i,n}, i=1,2$ are phonon annihilation operators, and
 the Pauli matrices $\sigma_{ln}$ represent two-level electron system.
 They satisfy identities
$\ [\sigma_{ln},\sigma_{jn}]=i\sigma_{kn},
 \ l=x,y,z$, representing $1/2$-pseudo-spins related to
 the electron densities in a usual way, i.e.
$\sigma_{xn}=\frac{1}{2}(c^{\dag}_{1,n}c_{2,n}+
c^{\dag}_{2,n}c_{1,n}),\ $
$\sigma_{yn}=\frac{1}{2i}(c^{\dag}_{1,n}c_{2,n}-
c^{\dag}_{2,n}c_{1,n}),  $\quad
$\sigma_{zn}=\frac{1}{2}(c^{\dag}_{1,n}c_{1,n}-
c^{\dag}_{2,n}c_{2,n}),$ $ \  I_n= \frac{1}{2} (c^{\dag}_{1,n}c_{1,n}+
c^{\dag}_{2,n}c_{2,n}) $ is a unit matrix, and
$c_{j,n}$ are electron annihilation operators.
The operator $R_{\pm 1,j}=e^{\pm ipa}$ of the displacement by a
lattice constant $\pm a$ acts in both the electron and phonon space,
$R_{\pm 1,j}f_n=f_{n\pm 1} R_{\pm 1,j} $.

In terms of the creation-annihilation electron and phonon operators
the Hamiltonian can be cast as follows:

 \begin{eqnarray}
H=\sum_{n} [\Omega\sum_{i=1,2} (b_{i n}^{\dag}b_{i n}
+\frac{1}{2} ) +\frac{\alpha}{2}(n_{1n}-n_{2n} )(b_{1n}^{\dag}+b_{1n})
\nonumber\\
-\frac{\beta}{2} (c_{1n}^{\dag}c_{2n}+c_{2n}^{\dag}c_{1n}
)(b_{2n}^{\dag}+b_{2n})\qquad \nonumber\\
-\frac{T}{2} \sum_{j=1,2}( c_{j,n}^{\dag}c_{j,n+1}+H.c. )]\, .\qquad
\label{1}
\end{eqnarray}

For  $\beta=-\alpha $, the  interaction part of (\ref{Hams1})
\begin{equation}
\alpha \left (\matrix{ b_{1n}^{\dag}+b_{1n}, \ b_{2n}^{\dag}+ b_{2n}\cr
 b_{2n}^{\dag}+b_{2n} ,\  -( b_{1n}^{\dag}+b_{1n})   }\right)
 \label{sym}
\end{equation}
yields the rotationally symmetric $E\otimes e$ form  \cite{Obrien:1993}
with a pair (an antisymmetric and a symmetric under reflection) of
double degenerated vibrations. This is, e.g., the case of $Cu^{++}$
ions with $d^9$ configurations in high-$T_c$
cuprates\cite{Bednorz:1986}$^{,}$\cite{Obrien:1993}.

 Taking $\alpha\neq \beta $ removes the
 degeneration of the vibronic states breaking the
 rotational symmetry of the electron-phonon interactions, the
 model still staying within the class
 of JT models \cite{Kaplan:1995}$^{,}$ \cite{Obrien:1993}$^{,}$
 \cite{Wagner:1984}.

  The dispersionless optical phonon mode $b_{1}$ splits
the degenerated unperturbed electron level ($j=1,2$)
while the mode $b_{2}$ mediates the electron
transitions between the levels. This latter term represents phonon-assisted
tunneling, a mechanism of the nonclassical (nonadiabatic) nature as well as
 is the pure tunneling in related exciton and dimer models.

 Evidently, Hamiltonian (\ref{Hams1}) ($\alpha\neq \beta$)
 is reflection-symmetric,  $G^{(el)} G^{(ph)}H=H $,

 \begin{eqnarray}
|2\rangle &=& G^{(el)} |1\rangle, \ (G^{(el)})^ 2=1,\nonumber\\
 G^{(ph)}_{1n} (b_{1n}^{\dag}\pm b_{1n}) &=&
- (b_{1n}^{\dag}\pm b_{1n}) G^{(ph)}_{1n}, \  (G^{(ph)}_{1n})^ 2=1,
\label{Gn}
\end{eqnarray}
where 
$G^{(ph)}_{1n}= \exp (i\pi b^{\dag}_{1n}b_{1n})$ is the phonon
reflection operator. While the phonon $1$ is antisymmetric under the
reflection, phonon $2$ remains symmetric.

In addition, the transfer part of (\ref{1}) exhibits $SU(2)$ symmetry
of the left- and right-moving electrons (holes).

Let us note that the quantum phonon assistance of the
electron tunneling ($\beta$-term in (\ref{1}) and (\ref{Hams1}))
constitutes the difference of the model from
the related dimer and exciton quantum models where instead of $\beta
\sum \limits _n (b_{2n}^{\dag}+b_{2n})\sigma_{xn}$ of (\ref{Hams1})
there stands $\Delta\sigma_{xn}$, where $\Delta $
is the distance between the levels
\cite{Sander:1973}$^{,}$\cite{Sonnek:1994}.

The local part of (\ref{Hams1}) can be diagonalized
in electron subspace by the Fulton-Gouterman  unitary operator
\cite{Fulton:1961} $U_n\equiv U_{2,n}U_{1,n}$, where
\begin{equation}
 U_{i,n}= \frac{1}{\sqrt 2} \left ( \matrix{1\ , \ G_{i,n}\cr  1\ ,
\ -G_{i,n}}\right ), \quad G_{i,n} =\exp (i\pi
b_{in}^{\dag}b_{in})\equiv G^{(ph)}_{i,n},
\label{Un}
\end{equation}
 as  follows
 \begin{eqnarray}
\tilde H_L
=\sum\limits_n  U_n H_{0n}U_n^{-1}=
\Omega\sum\limits_{n, i=1,2}
\left (b_{in}^{\dag}b_{in} +\frac{1}{2}\right )\nonumber\\
+\frac{1}{2}\sum\limits_{n}[\alpha (b_{1n}^{\dag}+b_{1,n})
-\beta (b_{2n}^{\dag}+b_{2n})G_{1,n})]I_n .
\label{diagh1}
\end{eqnarray}
On the other hand, in the transfer term
\begin{equation}
\tilde H_T = -\frac{T}{2} \sum_{n}\left (V_{n,1}R_1 + V_{n, -1}R_{-1}\right )
\label{diagh2}
\end{equation}
there appears a nondiagonality
\begin{eqnarray}
V_{n,\pm 1}= \left [(1+G_{1n}G_{1n \pm 1})I_n+
(1-G_{1n}G_{1,n\pm 1})G_{2n}\sigma_{zn}\right ]\nonumber\\
\times\left [(1+G_{2n}G_{2n \pm 1})I_n+ (1-G_{2n}G_{2,n\pm 1})
\sigma_{xn}\right ]. \qquad
\label{V}
\end{eqnarray}
Here, Pauli matrices transform as 
$U_{i} \sigma_{x}U_{i}^{-1}=  G_{i}\sigma_z$, $U_{i} \sigma_z U_{i}^{-1}=
\sigma_x$, and $ U_{i} (b_i^{\dag}+ b_i) U^{-1}_i = (b_i^{\dag}+b_i) 2\sigma_x$.

The diagonal terms of (\ref{V}) represent the polaron transfer within
one level while the off-diagonal ones represent the interlevel polaron transfer
through the lattice. Evidently, the
contribution of the off-diagonal terms proportional to $1-G_{in}G_{i,n+1}$ is
much smaller when compared with those proportional to
$1+G_{in} G_{i,n+1}$.

Because of nonconservation of the number of
coherent phonons, they are able even in
the ground state to assist electron transitions between the levels.
In the Hamiltonian \ref{diagh1},
the operator  $G_{1n} =(-1)^{b_{1n}^{\dag}b_{1n}}$ (\ref{Un}),
highly nonlinear in the phonon-$1$ appears mediated by phonons $2$.
It introduces multiple electron oscillations between the
split levels mediated by {\it continuous virtual absorption and emission of the
 phonons $1$}.
The effect is analogous  to Rabi oscillations in quantum optics
due to photons \cite{Rabi:1936}.
Let us note that Rabi oscillations assist  both the interlevel
onsite and intersite electron transitions mediated by the
electron transfer $T$.

For various local electron-phonon models, a number of phonon
variational wave functions using coherent and squeezed coherent
phonons  was proposed in order to reach the
best ground state \cite{Lo:1991}$^{,}$  \cite{Chen:1989}$^{,}$
\cite{Zheng:1990}$^{,}$ \cite{Feinberg:1990}.

Specially, for two-level systems
Shore et al. \cite{Sander:1973} proposed variational eigenfunctions
in the form of symmetric and antisymmetric combinations of the
 reflection symmetric components related to both the levels.
Each of these components was chosen in a form which accounted for mixing
 with the reflected state via new variational parameter. This choice was
 stated to be the best variational wave function, i.e. yielding the lowest
 ground state energy.
 The method was further developed by Wagner et al. \cite{Sonnek:1994}  for
 dimers and excitons. We shall adopt this approach in what follows by
 taking
\begin{eqnarray}
\Phi_{1,2}%([\gamma_{1q}(n),\gamma_{2q}(n), r_q(n)],\eta; [f(n),g(n)])\nonumber\\
\equiv \phi_1\psi_1 \pm\phi_2 \psi_2 \nonumber\\
= \frac{1}{\sqrt {C}}\left [(\phi_+ + \eta\phi_-)\psi_1\pm
(\phi_-+\eta\phi_+)\psi_2 \right ]\nonumber\\
=\frac{1}{\sqrt {C}}\left [\phi_+
(\psi_1\pm  \eta \psi_2)\pm\phi_-(\psi_2\pm \eta \psi_1)\right ],
\label{16}
\end{eqnarray}
with the upper sign for the ground state $\Phi_1$ and lower sign
for the excited state $\Phi_2$.
%Dependence of the phonon
%wave functions on variational parameters is as follows:
%$\phi_i\equiv \phi_i([\gamma_{1m}(n),\gamma_{2m}(n),
%r_m(n)],\eta). $\\

In (\ref{16}), $\phi_1= \phi_++\eta\phi_-$  and $\phi_2=
\phi_-+\eta\phi_+$ are phonon wave functions related to two levels with
mixing of the bare $\phi_+$ ($\phi_-$) and reflected $\phi_-$ ($\phi_+$)
parts mediated by the variational parameter $\eta$.

Further, $\psi_1$ and $\psi_2$ are components of the electron
state vector related to the upper and lower level, respectively,
\begin{equation}
\psi_{1}(n)= c^{\dag}_{1n}|0_{el} \rangle ,
\ \psi_{2}(n)= c^{\dag}_{2n}|0_{el}  \rangle ,
\label{f,g}
\end{equation}
where  $|0_{el}\rangle$ is the electron vacuum.
 Note that the last line of (\ref{16}) allows
us to interpret alternatively the parameter $\eta$ as reflection parameter
of the electron states as well.

 The squeezed coherent phonon wave functions
\begin{eqnarray}
\phi_{\pm}[\gamma_{1m}(n),\gamma_{2m}(n),r_{m}(n)]\quad\nonumber\\
\equiv
D_{1,\pm}[\gamma_{1m}(n)]D_{2}[\gamma_{2m}(n)] S_{1}[r_m(n)]|0_{ph}\rangle ,
\label{phi}
\end{eqnarray}
are related to lower ($\phi_+$) and higher ($\phi_-$) level;
$|0_{ph}\rangle$ is the phonon vacuum.

The generators of the coherent phonons are 
functionals of related displacements
\begin{eqnarray}
D_{1,\pm}[\gamma_{1m}(n)]&=&\exp \sum _{m}
\pm [\gamma_{1m} (n) b_{1m}^{\dag}-\gamma_{1m} ^*(n)
b_{1m} ],
\label{dis1}\\
D_{2}[\gamma_{2m}(n)]&=&\exp \sum _{m}
[\gamma_{2m}(n) b_{2m}^{\dag}-\gamma_{2m}^*(n) b_{2m} ],
\label{dis2}
 \end{eqnarray}
 as well as the generator of squeezing
 \begin{equation}
S_{1}[r_m(n)]= \exp \sum\limits_{m}
r_{m} (n) (b_{1m}^{\dag 2}- b_{1m}^2).
\label{sq}
\end{equation}
Respective generator of the mode $2$, $S_2 [r_{2m}(n)]= 1$,
because Hamiltonian (\ref{1}) is linear in $b_2, b_2^{\dag}$ and $r_{2m}(n)=0$.

The normalization condition then reads
\begin{equation}
\frac{1}{N^2} \sum\limits _{m,n } C_{1m}(n) = C,
 \label{C}
 \end{equation}
 where
\begin{equation}
C_{1m}(n)=1+\eta^2+2\eta \exp
 (-2|\tilde\gamma_{1m}(n)|^2),
 \label{C1}
 \end{equation}
and
\begin{equation}
\tilde\gamma_{1m}(n) \equiv \gamma_{1m}(n) \exp (-2r_{m}(n)).
\label{gamma}
\end{equation}

Generally, $\gamma_{jm}(n)$ is a nonlocal quantity which represents the
displacement of the mode $j$ at the site $m$ due to an electron at the
site $n$, $\gamma_{jm}(n)=
\frac{1}{\sqrt N}\sum \limits_q \gamma_{jq} (n)\exp (-iqma)$.
We take
\begin{equation}
\gamma_{j q}(n)= \frac{\gamma_{j}}{\sqrt N} \exp(iqna)\rightarrow
\gamma_{j m}(n)=\gamma_j \delta_{m,n},
\label{Q}
\end{equation}
where $\gamma_j $ are independent of $n$. Eq. (\ref{Q}) indicates
that the phonon displacement accompanies the electron at the site $n$.
The same is valid for the squeezing, $r_m (n)=
r\delta_{m,n}$.

Full Bloch solution to the transformed lattice Hamiltonian $\tilde H= \tilde
H_L+\tilde H_T $, (\ref{diagh1}), (\ref{diagh2}) is chosen as a
generalized Fulton-Gouterman variational ansatz
\cite{Fulton:1961}$^{,}$ \cite{Sonnek:1994} in the vector form (Eq.
(\ref{16}))
\begin{equation}
\Psi_{FG}(k) = \frac{1}{\sqrt {C}}\sum_n \exp (ikna)
\left(\matrix{\phi_1(n)\psi_{1}(n)\cr
\phi_2(n)\psi_{2}(n)}\right ) ,
\label{fuga}
\end{equation}
where the electron parts $\psi_i$ are defined by (\ref{f,g}) and the
phonon parts $\phi _i (n)$ by (\ref{16}) and (\ref{phi})-(\ref{sq}).

 We have investigated the model (\ref{1}) in a former paper
\cite{Majernikova:1996} with respect to the stability of a soliton ground
state against quantum fluctuations. However, the variational wave function
we used did not account for the two-center nature of the wave function
 here respected by the reflection parameter $\eta$ (\ref{16}).

In what follows the ground state energy as function of optimized
variational parameters $\eta$,
$\gamma_{1} ,\gamma_{2}, r $ will be determined.

\section{The ground state}
When considering the ground state of $\tilde H=\tilde H_L +\tilde H_T$,
(\ref{diagh1}) and (\ref{diagh2}), we omit the electron
and phonon dynamic terms by setting $ k=0, q=0 $.

By averaging $\tilde H= \tilde H_L+ \tilde H_{T}$ ((\ref{diagh1}),
(\ref{diagh2})) over the  phonon wave functions
(\ref{fuga}) with (\ref{16}) and (\ref{phi}) one obtains for the site
Hamiltonian (\ref{diagh1}) (Appendix B)
 \begin{equation}
\frac{\langle \tilde H_L\rangle}{N}\equiv H_{ph} + H_{int} ,
\label{HD}
 \end{equation}
 where
\begin{eqnarray}
 H_{ph} = \frac{\Omega}{2}(\cosh 4r +1) I  \label{hav1}\\
+ \Omega \frac{1+\eta^2-2\eta e^{-8r_1}\exp(-2|\tilde \gamma_{1}|^2)}
{1+\eta^2+2\eta\exp(-2|\tilde \gamma_{1}|^2)} |\gamma_{1}|^2 I
+ \Omega |\gamma_{2 }|^2 I,
\label{hav2}
\end{eqnarray}

\begin{eqnarray}
 H_{int} \equiv
H_{\alpha}+H_{\beta} =
 \frac{\alpha (1-\eta^2)}
 {1+\eta^2+2\eta\exp(-2|\tilde \gamma_{1}|^2)}\gamma_{1}\sigma _{z}\nonumber\\
-\frac{\beta [(1+\eta^2)\exp(-2|\tilde \gamma_{1q}|^2)+2\eta]}
{1+\eta^2+2\eta\exp(-2|\tilde \gamma_{1}|^2)}
\gamma_{2} I  ,
\label{hav12}
\end{eqnarray}
where $H_{\alpha}$ and $H_{\beta}$ are $\alpha$- and $\beta$-dependent parts
of the interaction $H_{int}$ (\ref{hav12}).
From the transfer Hamiltonian $\tilde H_T$ (\ref{diagh2}) there remains
\begin{eqnarray}
\frac{\langle\tilde H_T\rangle }{N}=-T  \exp(-W )M  \nonumber\\
\equiv 2(H_{T1}I+H_{T2}i \sigma_{y}+ H_{T3}\sigma_{z} + H_{T4}\sigma_x
).
\label{havt}
\end{eqnarray}
The Debye-Waller factor $\exp (-W) M $ in (\ref{havt})
is given in the Appendix B. The differential part of the Debye-Waller
factor $W$ in the ground state is zero. For $M $ it holds               
\begin{eqnarray}
M = \left [( 1+E_{1}^2 )I +
(1-E_1^2) E_2 \sigma_{zn} \right ]\nonumber\\
\times \left [(1+E_2^2) I + (1-E_2^2)\sigma_{x} \right ],
\label{T2}
\end{eqnarray}
where  $E_i$ are given by (Appendix B)
\begin{equation}
  E_1 \equiv \exp\left (-2|\tilde\gamma_{1}|^2\right ), \
   E_2 \equiv \exp\left (-2|\gamma_{2}|^2\right ).
   \label{Ei}
\end{equation}
Expressions $E_i$ (\ref{Ei}) in the ground state are independent of
$q$ and $n$ because of the form of $\gamma_{iq}(n)=
\gamma_i \exp (-iqna)$ (\ref{Q}).
This substantially simplifies subsequent calculations leaving
$E_{i}$ and $ C_{1}$ as functions of $\gamma_i^2 $ independent on $n$.
The transfer terms $H_{Ti}, \
i=1,2,3,4$ in (\ref{havt}) with (\ref{T2}) are expressed as
\begin{eqnarray}
H_{T1}= -\frac{T}{4} (1+E_1^2)(1+E_2^2),
\nonumber\\
H_{T2}= -\frac{T}{4}(1-E_1^2)(1-E_2^2)E_2,
\nonumber\\
H_{T3}= -\frac{T}{4} (1-E_1^2)(1+E_2^2)E_2,
\nonumber\\
H_{T4}= -\frac{T}{4} (1+E_1^2)(1-E_2^2).
\label{hti}
\end{eqnarray}
  Here, $H_{T1}, H_{T3}$ and $H_{T2}, H_{T4}$  are
diagonal and off-diagonal terms of (\ref{havt}), respectively.

The effective polaron potential in (\ref{HD}) (with
 (\ref{hav1})-(\ref{Ei})) as a highly nonlinear function of
 $\gamma_1$ and $\gamma_2$ is visualized in Fig. 1.

The potential exhibits two sets of minima related to two competing
ground states:\\
(i)
Two nonequivalent broad minima related to both the levels
(\ref{dis1}) at $\pm \gamma_1 \neq 0$ and $\gamma_2$ close to $0$; \\
(ii) One narrow minimum at $\gamma_1$ close to $0$, $\gamma_2 >0$, where
both levels approach close together. This minimum develops at 
growing $\beta$; evidently its behaviour depends also on $T$ value
because of nonlinearity of the Debye-Waller factor $M$ ((\ref{T2}) and
(\ref{Ei})).

The ground state energies related to these two sets of competing minima
were calculated numerically.
%Variational parameters $\gamma_1,\ \gamma_2,\ r,\ \eta$
%are coupled in a highly  nonlinear way by a set of transcendent algebraic
% equations.
The result of the numerical minimalization of the diagonalized form of energy
(\ref{HD})
\begin{eqnarray}
E=  H_{ph}+H_{\beta }+ H_{T_1}-\left [(H_{\alpha}+H_{T_3})^2+
H^2_{T_4}-H^2_{T_2}\right ]^{1/2}
\label{EF}
\end{eqnarray}

 can be written symbolically as
\begin{equation}
E_G (\mu, \chi,\Omega/T)= E (\{\gamma_{1q}, \gamma_{2q}, r, \eta
\}_{min})/T.
\label{HG}
\end{equation}
Here, the index $min$ denotes the optimized values.
The model parameters
\begin{equation}
\mu=\frac{\alpha^2}{2\Omega T},\quad
\chi=\frac{\beta}{\alpha}, \quad \frac{\Omega}{T}
\label{mu}
\end{equation}
are parameters of the effective
interaction, asymmetry and nonadiabaticity, respectively. Energy $E_G$
(\ref{HG}) is in scaled units, renormalized by $T$. The results of the
numerical evaluations of the ground state energy (\ref{HG}) are depicted
in Figs. 2a, 2b  different model parameters.

 One can distinguish there two regions depending on the value of $\chi$ with different behaviour of ground state in each of them:\\
(i) The ground state pertaining to the lower broad minimum at $\gamma_1 <0$
 and $\gamma_2\approx 0$ with a small reflection part at $\gamma_1>0$
 is referred to as a "heavy" region.
It corresponds to predominantly intralevel "heavy" polaron which is
represented by a two-peak wave function,
both peaks representing a harmonic oscillator ($\sim \exp
[-(x\mp \gamma_1)^2]$) displaced by the value $\pm \gamma_1$ (as it is seen
from the form of the ansatz (\ref{dis1})). The "heavy" polaron
is dominant at $\chi<1$ where the broad minimum  $\gamma_1<0$ is dominating.\\
(ii) Close to $\chi=1$, the
energies of two minima go very close together (their difference is
of the order of the phonon energy $\Omega$). They
drop to one narrow minimum which represents a new ground state.
{\it Close to $\chi=1$, continuous transition to a new ground state occurs}.
It stabilizes at $\chi>1$, where (optimized) $\gamma_1$ value is close to $0$ and $\gamma_2>0$.
 This region is referred to as a "light" polaron region because, owing
 to the abrupt decrease of $\gamma_1\approx 0$ at $\chi>1$, the
 effective mass of the intralevel polaron drops to almost its
 free-electron value, i.e., {\it the polaron selflocalization vanishes
 in the "light" region}.
 Because of this "undressing", the transport characteristics of the excited
 electron would increase. Moreover, due to the tiny distance between the
 levels, their coupling takes place by the exchange of virtual phonons $1$.
  At suitable conditions, in the excited state, when both
 levels are occupied by electrons of opposite spins,
 {\it the mechanism of virtual phonon exchange implies
 the pairing of electrons, i.e. formation of "light" bipolarons }.

 The ground state energy,
 especially its behavior dependent on pairs of parameters $\chi$,
 $\Omega$ and  $\chi$, $\mu$ is illustrated in details in
 Fig. 2. While being weakly $\Omega$-dependent, the energy  strongly
decreases with $\chi$ inside the "light" phase (Fig. 2a).
 The position of the phase line is slightly shifted from $\chi=1$ at
$\Omega=0$ to higher values of
$\chi$ with increasing $\Omega$.  This is consistent with
 the fact that the phonon fluctuations are most effective
 when the difference of the energies of the phases is of the order
 of the phonon energy.

The ground state energy in the "heavy" region ($\chi<1$) is
 independent of $\chi$, its decrease inside the "light" region
($\chi>1$) is dependent on the effective coupling $\mu$ (Fig. 2b).
The energy decrease due to $\mu$ is stronger in the "light" phase
(depending on $\chi$) than in the "heavy" phase.

 The region of importance of reflection measure $\eta $  and $r$ is
 illustrated in the
 Fig. 3. There we show the difference between the ground state energy
 with $r$ and $\eta$ omitted and ground state with four variational
 parameters calculated numerically. However,  contribution due to $r$
 $E_{G}(\eta, r=0)-E_G \sim 2.10^{-3}$ of the maximum value in Fig. 3.
 The said parameters
  turn out to be most relevant in E$\otimes$e JT case ($\chi=1$).

The effects of $\eta $ and $r$ on the ground state is apparent
only at moderate couplings in the "heavy" region,  reaching their
maximum at $\chi=1$.
 The narrow minimum (the "light" region) is resistant against $\eta$.

In order to demonstrate important properties of displacements ($\mu $
and $\Omega$ dependence) it is sufficient, except of the region of
importance of the reflection (Fig. 3) inside the "heavy" region
apparent for $\gamma_2$ (Fig. 4),
to calculate them in the limit $\eta=0$.
We shall use Eq. (\ref{EF}) approximated for the cases of both "heavy" and
"light" polaron and obtain implicit expressions for $\gamma_i$ which however
are good to visualize their behaviour in both regions:

(i) "heavy" polaron ($\gamma_2$ small, $ E_2\approx 1,  H_{T_2},
H_{T_4}\approx 0$):
\begin{eqnarray}
\gamma_{1} & \approx &\frac{-\alpha e^{4r} }
{2T E_1^2\left
[2+\frac{\Omega e^{4r}}{T E_1^2} \right ]},
%\frac{\beta}{2T E_2^2} \frac{E_1
%}{\left (1+E_1^2+\frac{\Omega}{T E_2^2}\right)};
\label{ie1}
\end{eqnarray}
$\gamma_2 $ is small except of the region of fluctuations
visualized in Fig. 4. One can see, that
the selflocalization due to phonons $2 $ in the "heavy" region
$\sim - \gamma_2^2$ clearly implies the "cave"  in the ground
state energy due to the reflection measure $\eta$ (anharmonicity) of the
ground state (Fig. 3).

(ii) "light" polaron ($\gamma_1$ small, $ E_1\approx 1,
H_{T_2}, H_{T_3}\approx 0$):
\begin{eqnarray}
\gamma_{1}& \approx & \frac{-\alpha  e^{4r} }{2 T E_1^2
\left (1+\frac{\beta^2 }{2\Omega T}+  \frac{\Omega e^{4r}}{TE_1^2}\right
)}\approx 10^{-8},
\nonumber\\
\gamma_{2}& \approx &\frac{\beta}
{2TE_2^2(2+\frac{\Omega}{TE_2^2})}.
\label{ie2}
\end{eqnarray}
In the "light" region the fluctuations of $\gamma_1$ are missing as well as
the fluctuations of the energy. This is consistent with the above result
of the resistance of the narrow minimum against $\eta$.

For both "heavy" and  "light" polaron a dependence of
$\gamma_{i}$ on the nonadiabaticity parameter $\Omega/T$ appears.
It implies the dependence of
the Debye-Waller factor and consequently of the polaron mass
on the phonon frequency $\Omega$. This can be thought of as an analogy
of the {\it isotope effect} at zero temperature.

In the "heavy" phase, electron transitions mediated by phonons $2$ to the
upper level enhance fluctuations $\sim \gamma_2$ which mix phonons $2$ with
phonons $1$ and contribute to the fluctuations of the ground state
energy of the heavy region. This is the reason for the similarity of
the results in Figs. 3 and 4.

\section{Metal-insulator transition}

To illustrate the effect of various interactions let us consider
first a simple two level case ($T=0$).
The ground state will be obtained by the numerical minimalization of
energy
\begin{equation}
E^{0}= H_{ph}+H_{\alpha}+H_{\beta },
\label{E-0}
\end{equation}
where $H_{\alpha}$ and  $H_{\beta}$ are $\alpha$ and $\beta$
dependent components of the
interaction term $H_{int}$ (\ref{hav12}).
The result is of the form
\begin{equation}
E^0_{G}= \frac{\Omega}{2}(\cosh 4r^0+1)-H_{P\alpha} -H_{P\beta},
\label{E0}
\end{equation}
where the label $0$ specifies the optimized variational values.
In (\ref{E0}), $H_{P\alpha}$ and $H_{P\beta}$ are contributions of the
selflocalization energies of two coupled polarons proportional
to $\alpha^2/2\Omega$ and $ \beta^2/2 \Omega$, respectively.
 The splitting of the levels is implied to be
\begin{equation}
\Delta^{0}= 2|E^{0}_G|.
\label{spl}
\end{equation}

Respectively a lattice of JT molecules ($T\neq 0$) shows
 either metallic or insulating phase behavior depending on whether the
 splitted bands overlap or not. Usually the resulting  situation depends
 strongly on the phonon renormalization of the gap and of the electron
 band including phonon fluctuations.

Our aim here is to investigate the metal-insulator transition due to
the competition of the polaron localization and the
delocalization due to the transfer including the phonon assisted electron
transfer effects. Namely, the transport terms $H_{Ti}$ compete the
level splitting (\ref{spl}) because of the complex interplay of the diagonal and
off-diagonal contributions to the band width. In order to find the value of the gap as an order parameter of the
metal-insulator phase transition it is necessary to perform usual
renormalization:
 $SU(2)$ symmetry of the right- and left-moving electrons
and holes of the transfer part of Hamiltonian (\ref{Hams1})
makes it possible to
 split electron operators onto an electron and a hole part,
\begin{equation}
c_i=c_{i,+}+c_{i,-},  \ i=1,2,
\label{Psi}
\end{equation}
the signs $\pm$ indicating directions of the electron motion. 
In view of the reflection symmetry of the two-level local part of
Hamiltonian  (\ref{Hams1}) against the level $E=0$ in the middle between
the levels electron operators (\ref{Psi}) can be rewritten in a
renormalized form
\begin{equation}
c_1=c_{1,+}^{(el)}+c_{1,+}^{(h)\dag},\quad
c_2= c_{2,-}^{(h)\dag}+c_{2,-}^{(el)}.
\label{ren}
\end{equation}
This renormalization implies the change of the sign of $T$ ($H_{T_i}$)
for holes $(E<0)$.

There are two pairs of electrons and holes,
$c _{1,+}^{(el)}$,  $c_{2, -}^{(h)\dag}$ and
$c _{2,-}^{(e)}$, $c_{1,+}^{(h)\dag}$, i.e., two state vectors
$\left ( \matrix{f_+\cr g_-^*}\right)$ and $\left ( \matrix{-g_-\cr
f^*_+}\right)$   which satisfy equations
\begin{equation}
H_m \left(\matrix{ f_+\cr g_-^* }\right ) +
\left(\matrix{-\Delta/2,\ 0\cr 0, \ \Delta/2}\right ) \left(\matrix{
-g_-^*\cr f_+} \right )=0,
\end{equation}
\begin{equation}
 H_m \left(\matrix{ -g_-\cr f^*_+ }\right ) + \left(\matrix
{\Delta/2,\ 0\cr \ 0, \ -\Delta/2}\right )\left(\matrix{ f_+^*\cr g_-}
\right )=0,
\label{ee1}
\end{equation}
where $H_m$ results from the minimalization of (\ref{HD}) with respect to the
variational parameters, (\ref{HG}),

\begin{equation}
H_m=\left
(\matrix{-E_G, \ 0\cr 0 , \ E_G}\right ),
\label{H}
\end{equation}
and $E_G$ given by (\ref{HG}).

The solution for the gap $\Delta$ reads finally (see Figs. 2b-4b)
\begin{equation}
\Delta=2 |E_G(T\rightarrow -T)|
\label{gap}
\end{equation}
and implies a condition for the stability of either the insulating phase,
if $E_G(T\rightarrow -T) > 0$, or the metallic phase in the opposite case,
where $\Delta\equiv 0$.

Fig. 5. depicts both I-M  and "heavy"-"light" transitions.
In the "heavy" phase, at small $\Omega$ the I-M transition line is
determined by the critical coupling  $\mu_H (\Omega)$ independent of
$\chi$ up to $\chi=1$. In the "light" phase, the critical coupling
$\mu_L (\Omega, \chi) < \mu_H(\Omega)$ decreases with increasing $\chi$.
The point $(\chi=1, \ \mu_H (\Omega))$ is the only point of coincidence of the
"heavy"-"light" transition and the I-M transition.  At small $\mu$ the
fluctuations support the metallic phase in the "light" region (Fig. 5a),
while at small $\chi$ (and sufficiently large $\mu$ that the insulating
phase could be stable) they support the metallic phase in the heavy region
(Figs. 5a, b).

%%%%%%%%%%%%%%%%%%%%%%%%%%%%%%%%%%%%%%%%%%%%%
The I-M transition can appear in the "heavy" region  at large
$\mu$ (\ref{mu}) and small $\Omega$  (Fig. 2a). From Fig. 3 (medium values
of $\mu$)
one can locate the region of relevant reflection $\eta$ into the metallic phase,
where the bands overlap. The "light" region ($\chi>1$) is resistant
against $\eta$: no similar effect as that one shown in Fig. 3 there exists.
Therefore, in the "heavy" region the metallic
phase is strongly supported due to the fluctuations at weak couplings
$\mu$. "Heavy" phase is therefore metallic in the broad range of
parameters except of strong couplings (Fig. 2b).
At certain  (large) $\mu$ the I-M transition drops close
to the line $\chi=1$, so
that the transition coincides with the "heavy" to "light" polaron
transition.
With decreasing $\mu$ the I-M transition line moves to larger $\chi$
 (Fig. 5b): the M-I transition moves away from
the "heavy"-"light" transition into the "light" region.
 In the "light" region the I-M transition line moves to smaller $\mu$ when
 compared with the heavy region (Fig. 6).

As one can also see from Figs. 6 and 8, the I-M transition
line in the phase diagram $\mu, \Omega $ is strongly $\Omega $-dependent.
 The phase line in the plane ($\mu, \Omega$) between the phases shows the broadening of
the metallic phase with growing $\Omega$, i.e. the metallic phase is supported
by the quantum fluctuations.
Respective ground state energy slightly increases with $\Omega$.
%%%%%%%%%%%%%%%%%%%%%%%%%%%%%

To make effects of different contributions $H_{T_i}$ more transparent,
we can first calculate analytically effect of the diagonal terms
 $H_{T_1}$ and $H_{T_3}$.
 (Let us note that neglecting  the off-diagonal transfer,
 i.e.   $H_{T_i}, \ i=2,4$
in (\ref{hti}), means neglecting the additional splitting of the bands.
More precisely, it means neglecting the off-diagonal terms in Eq. (\ref{V}) ).

 Minimalization of the energy (\ref{EF}) with $H_{T_2}=H_{T_4}=0$
\begin{equation}
E^1 = H_{ph}+ H_{\alpha}+H_{\beta }+H_{T_1}+H_{T_3}
\label{T13}
\end{equation}
 implies the ground state of a similar structure
as that one of (\ref{E0}),
\begin{equation}
E^{1}_G= \frac{\Omega}{2}(\cosh 4r^1+1) -H^1_{P\alpha}
-H^1_{P\beta}+H^1_{T_1}+H^1_{T_3},
\end{equation}
where the label $1$ specifies the optimized variational values.
Respective splitting yields
\begin{eqnarray}
\Delta^{1}= 2|E_G^1- 2(H^1_{T_1}+H^1_{T_3})|\nonumber\\
=2| H^1_{\Omega}-H^1_{P\alpha}-H^1_{P\beta}-H^1_{T_1}-H^1_{T_3}|.
\label{D1}
\end{eqnarray}

The qualitative picture of the
role of different contributions in the formation of the gap is given
in Fig. 7.

The exact phase diagram is compared with the result of (\ref{D1})
in Fig. 8.
Here one can see  that the main contribution to the shift of the phase
diagram in support of the
metallic phase is on account of the off-diagonal electron transfer terms
 $H_{T_2}$ and $H_{T_4}$  (\ref{hti}) related to phonon assisted Rabi
 oscillations. These contributions play important role in the "light" phase
 (Fig. 2b).

\section{Squeezed quantum ground state (vacuum) fluctuations of phonons}

In Fig. 9 we display numerical results for the coordinate and
momentum fluctuations of the phonon in the "heavy" region ($\chi<1$) of
the ground state. Evident
anharmonicity peaks in the region of relevance of the reflection $\eta$
(Fig. 10) are due to the mixing with the phonon of the higher level.
The peaks are related to the ground state
(squeezed vacuum) fluctuations of the phonon coordinate
$Q_1=\frac{1}{\sqrt{2}} (b_1^{\dag}+b_1)$
\begin{eqnarray}
\Delta^2 Q_1= \langle Q_1^2\rangle-\langle Q_1\rangle ^2=\qquad \nonumber\\
=\frac{1}{2}\exp(4r)
 \Bigl [1+4\tilde\gamma_{1}^2-\frac
{4(1-\eta^2)^2\tilde \gamma_{1}^2}{(1+\eta^2+2\eta
\exp(-2\tilde\gamma_1^2))^2}\Bigr ],
\label{unc1}
\end{eqnarray}
$\tilde \gamma_1$ defined by (\ref{gamma}). For the variational parameters
$\gamma_1, \eta, r $ we have to take their optimized (ground state) values.
For the related momentum $P_1=\frac{i}{\sqrt{2}} (b_1^{\dag}-b_1)$,
one obtains the ground state fluctuation
\begin{equation}
\Delta^2 P_1= \langle P_1^2\rangle-\langle P_1\rangle ^2=
\frac{1}{2}\exp(-4r)
\label{unc2}
\end{equation}
 where again the optimized value for $r$ is meant.

Their product $\Delta^2 Q_1\Delta^2 P_1$ for
the oscillator in the ground state (Fig. 8) exhibiting a peak in the
region of relevant reflection $\eta$ is the evidence for its anharmonicity
due to the coupling with the higher level oscillator. The reflection variational
parameter $\eta$ is efficient only in the "heavy" region ($\chi<1$),
for $\chi=1$ the fluctuations decrease close towards the classical value
$0.25$ of the harmonic oscillator near the narrow minimum:
the stable ground state of
the "light" polaron is achieved at $\chi=1$.
In the "light"  region the oscillator remains close
to harmonic one;  no analogous effects to those in the "heavy" region
 occur. The results displayed on Fig.8 stay in evident correlation with the
 region of relevance of the reflection
parameter $\eta$  of the Fig. 3 (see its detail in Fig. 10 for the
corresponding region of parameters). Similar effects of
anharmonicity were found by numerical
simulations in the related model by Borghi et al. \cite{Borghi:1995}
mentioned in the Introduction.

\section{Conclusions}

The selfconsistent polaron potential provided by the Holstein intralevel
and interlevel electron-phonon couplings in our two-level lattice model is
highly nonlinear function of phonon displacements (Fig. 1).
As a consequence, there occurs a competition
between the regime of two nonequivalent broad minima at $\pm \gamma_{1}$ and
$\gamma_2$ close to $0$ related to two electron levels and the
regime of one narrow minimum at $\gamma_{2}>0$ and
$ \gamma_1$ close to $0$ when both broad minima collapse to a single one,
so that both levels drop to that minimum. The broad
minima dominate at $\chi<1$, the narrow one at $\chi>1$. At
E$\otimes$e Jahn-Teller symmetry ($\chi=1$) energies of the broad and narrow
minima coincide and the phonon-assisted tunneling due to the
nonadiabatic fluctuations reaches its maximum close to this limit.
When $\chi $ is approaching to $1$, there occurs pairing of the levels due to
the interlevel onsite and intersite polaron
tunneling mediated by the exchange of virtual phonons.
The narrow minimum is suppressed if the electron transfer energy $T$
decreases.

Hence, we have identified two regions of stability: (i) the region of
the dominating {\it "heavy" polaron} ($\chi<1$) related to the broad minimum at
negative $\gamma_1$ (the absolute minimum)
and $\gamma_2\approx 0$ and (ii) the region of the dominating
{\it "light" polaron} ($\chi>1$) related to the narrow minimum at
$\gamma_1\approx 0$ (Fig. 1).
The almost {\it vanishing polaron selflocalization} in the "light region"
("undressing") means {\it enhancement of the polaron transfer} due to
respective decrease of the polaron effective mass.
Even in the "heavy" phase the effect can be present due to fluctuations
of the "light" phase (Figs. 3 and 4).

Close to the transition from "heavy" to "light" polaron at $\chi=1$,
a continuous transition to {\it new ground state that is stable at
$\chi>1$ occurs} (Fig. 2).
 In  this ground state the {\it formation of "light" bipolarons} is
possible which are characterized by the  enhancement of the effective
transfer due to the vanishing selflocalization.
The pairing of the polarons is caused by
 virtual exchange of phonons $1$ mediated by phonons $2$ between the electrons
in both levels at suitable configuration.
 This effect might be of interest from the point of view of bipolaron
 mechanisms of the superconductivity \cite{Mott:1995}.
 The I-M transition and "heavy"-"light" phase coincide
in the crossing point of the lines $\chi=1$ and $\mu_H (\Omega)$
(critical line of I-M transition in the "heavy" phase, Fig. 5a).

The Debye-Waller factor (\ref{T2}) shows nonadiabatic dependence on the
phonon frequency $\Omega$ through $\gamma_i$, (\ref{ie1}), (\ref{ie2}).
Then, the effective transfer parameter of electrons is respectively
increased and the electron effective mass decreased. This effect reminds
on the "isotope effect", however, at zero temperature.

 The metal-insulator transition occurs in both regions: "heavy" region
is the region of dominating metallic phase, where the transition to the
insulating phase can occur only at sufficiently large $\mu$ (Fig. 5). On the
other hand, in the "light" region
 the I-M transition can occur at relatively small $\mu$ (Figs. 5 and 6).
 For certain $\mu$ the {\it I-M transition can
coincide with the "heavy-light" polaron transition} at $\chi=1$ (Fig. 5a).
The interlevel and the combined interlevel and intersite electron
Rabi oscillations in the ground state,
are identified to be the main reason for the {\it shift of the I-M transition
line to stronger effective couplings,
i.e. in the support of the metallic phase} (Figs. 5 and 7).
Fig. 5a illustrates the existence of the "light" polaron in both the
insulating and metallic phase.

{\it Fluctuations of the phonon conjugated coordinates and their product
exhibit peaks} (Fig. 9) in the region of parameters
where the effect of $\eta$  (Figs. 3, 4)
is relevant. Consequently, the effect appears only {\it in
the "heavy" region} reaching its maximum at $\chi=1$ and growing with
$\Omega$. The peaks evidence for {\it the strongly anharmonic
behavior} of the "heavy" polaron in the region of relevance of $\eta$.
In the "light" region,
there occurs a degeneracy over $\eta$ (no effect of $\eta$ on the
minimum of the potential), i.e. no analogous effect of
the fluctuations, but nearly a harmonic oscillator is observed.
This is obviously due to the specific form of the variational ansatz,
which is inspired by the shape of the "pure" heavy polaron (i.e.
with $\chi<< 1$). Therefore the ansatz (\ref{16}), reflecting the
essentials of the  "heavy" polaron, is proper for describing the
heavy region, but when concerning the "light" region,
it presents, in some extent, a heuristic extrapolation.
We could have started from another ansatz
underlying the essentials of "light" polaron (which could be, for example,
inspired for the "pure" case of $\chi>>1$) in order to investigate the
"light" region and gain some extrapolation towards smaller $\chi$.
Naturally, also the transition in E$\otimes$e Jahn-Teller case
($\chi\simeq 1$), needs further considerations.

\section {Acknowledgements}
We thank Prof. M. Wagner for useful discussions and Prof. J. Pe\v rina
for careful reading of the manuscript.
The support from the Grant Agency of the Czech Republic
(Grant No. 202/01/1450) and partly also from the VEGA (Grant No.
2/7174/20) is highly acknowledged.
S. Sh. thanks the Department of Theoretical Physics of the
Faculty of Sciences, Palack\'y University in Olomouc for the
hospitality and Ing. I. Vik\'ar for partial support of his stay.\\

{\bf Appendix A.}\\

We shall use following formulas \cite{Yuen:1976}$^{,}$\cite{Kral:1990}
$$
D(\gamma)S(r)= S(r)D(\tilde\gamma), \  \tilde\gamma=\gamma e^{-2r}
\quad\eqno (A1)$$
$$S^{-1}(r)b S(r) = b\cosh 2r+ b^{\dag} \sinh 2r
\quad\eqno(A2) $$

$$ \langle 0 | S^{-1}(r_m(n))D^{-1}(\gamma_m (n))D_{m}(n+1)S
(r_m(n+1))$$
$$ = \exp \left ( -\frac{1}{2}[\tilde \gamma_m (n+1)-\tilde
\gamma (n)]^2\right ),
\quad\eqno(A3) $$

where $\tilde \gamma= \gamma e^{-2r}$.

For averages with using virtual Fock states $|n\rangle $ we use formulas
$$\langle 0| D^{-1}S^{-1} (-1)^{b^{\dag}b}SD|0\rangle \equiv {}_g
\langle\gamma |(-1)^{b^{\dag}b} |\gamma\rangle _g$$
$$=\sum\limits_{n=0}^{\infty} (-1) ^n|\langle n|\gamma\rangle|^2,
\quad\eqno(A4)$$
where $ |\gamma\rangle_g=SD|0\rangle =S|\gamma \rangle $ and,
$$\langle n|\gamma\rangle _g= \frac{1}{(n! \mu)^{1/2}} \left
(\frac{\nu}{2\mu}\right ) ^{n/2}H_n
\left(\frac{\gamma}{(2\mu\nu)^{1/2}}\right)$$
$$\times \exp \left
(-\frac{|\gamma|^2}{2}+\frac{\nu\gamma^2}{2\mu}\right ). \quad\eqno (A5)$$
Here, $\mu=cosh 2r, \ \nu=\sinh 2r$, $\mu^2-\nu^2=1$, here $\nu$ is
real.\\

{\bf Appendix B.}\\

The average of the Hamiltonian in (\ref{diagh1}) over the states
(\ref{fuga}) referred to one site results in

$\frac{\langle \tilde H_L\rangle}{N} =  H_{ph} + H_{int} $, where
$$ H_{ph} = \frac{\Omega}{2CN^2}\sum\limits_{m,n}C_{1n,m}(
\cosh(4r_m(n))+1) I_n$$
$$+ \frac{\Omega}{CN^2}\sum\limits_{m,n} (1+\eta^2-2e^{-8r_m(n)}
\eta \exp(-2|\tilde \gamma_{1m}(n)|^2)
)|\gamma_{1 m}(n)|^2  I_n$$
$$+ \frac{\Omega}{CN^2}\sum\limits_{n,m} C_{1m}(n)
|\gamma_{2 m}(n)|^2 I_n ,\quad\eqno(B1)$$
$$
 H_{int} \equiv H_{\alpha}+H_{\beta}
=\frac{1}{CN^2}\sum\limits_{n,m}
\Bigl\{\alpha (1-\eta^2)\gamma_{1 m}\sigma _{zn}$$
$$-\frac{\beta}{2} [(1+\eta^2)\exp(-2|\tilde \gamma_{1m}|^2)+2\eta]
\gamma_{2 m} I_n \Bigr \}
 +H.c. ,\quad\eqno(B2)$$
and
$$
H_T=-\frac{T}{CN^2}\sum\limits_{n,m}C_{1m}(n) \exp(-W_m(n))M_m (n) $$
$$\equiv 2(H_{T1}I+H_{T2}i \sigma_{y}+ H_{T3}\sigma_{z} + H_{T4}\sigma_x
). \quad\eqno(B3)$$

Calculation of the Debye-Waller factor $\exp(-W)M$ (\ref{havt}) and of
(\ref{Ei}) is based on the formulas (A3)-(A5). We approximated
 the average of a product in the transfer term by a product of averages.
In the continuum limit, from (A3) we receive for $W (n)$ in (B3)
$$
W(n)=\frac{1}{2N}\sum\limits_m\left ( \left|\frac
{d\tilde \gamma_{1m} (n)}{d n}
\right |^2 +  \left|\frac{d \gamma_{2m} (n)}{d n}
\right |^2  \right ).
\quad\eqno(B4)
$$
With the help of (A4) and (A5) one obtains for $E_m(n)=\langle
G_{1,m+1}G_{1m}\rangle $,
$$E_m(n)\equiv
\langle 0_{m+1}, 0_m | S^{-1}(r_{m+1}(n))S^{-1}
(r_{m}(n))$$
$$\times D^{-1}(\gamma _{1,m+1}(n))D^{-1}(\gamma_{1m}(n))
(-1)^{b^{\dag}_{1,m+1}b_{1,m+1}}$$
$$\times(-1)^{b^{\dag}_{1m}b_{1m}}
D(\gamma _{1,m}(n+1)) D(\gamma_{1,m+1}(n+1))
S(r_{m}(n+1))$$
$$S(r_{m+1}(n+1))|0_m, 0_{m+1}\rangle
 =\exp \left (-\frac{1}{2}\left [ |\tilde
 \gamma_{1m}(n)|^2+\right.\right.$$
 $$\left.\left.+ |\tilde \gamma_{1m}(n+1)|^2+
 |\tilde \gamma_{1,m+1}(n)|^2+ |\tilde \gamma_{1,m+1}(n+1)|^2
 \right.\right.$$
  $$\left.\left.
+\tilde \gamma_{1m}(n+1)  \tilde \gamma_{1m}(n)+
\tilde\gamma_{1,m+1}(n+1)\tilde \gamma_{1,m+1}(n)
\right.\right.$$
$$\left.\left.+
\tilde\gamma_{1m}(n)\tilde \gamma_{1,m+1}(n)
+\tilde\gamma_{1,m+1}(n+1)\tilde \gamma_{1m}(n+1)\right ]
 \right ).\quad\eqno(B5)$$
In the continuum limit (B5) yields
$E_m(n)\rightarrow \exp (-2 |\tilde \gamma_m(n)|^2)$.\\

$*$ Electronic address: majere@prfnw.upol.cz

$**$ on leave of absence from the Institute of Nuclear Research UAN,
Pr. Nauki 47, Kiev, Ukraine.

\newpage

\section{Figure Captions}

Fig. 1. Potential (\ref{HD}) for
 $\chi=\beta/\alpha=0.5, \ \mu=2$ and  $\Omega=0.05$.

Fig. 2. The ground state energy  (\ref{HG}) in the $\chi $-$\Omega $
plane at $\mu= 2.5$ (a) and
(b) in the $\chi$-$\mu$ plane at $\Omega=0.5$.
The "light" polaron is evidently much more
sensitive to effects of $\chi$ and $\mu$ than the "heavy" polaron. 

Fig. 3. The range of relevance of the reflection parameter $\eta$:
Difference  $\Delta E_G=E_G(\eta=0, r=0)-E_G$, for $\chi=1$,
 $E_G $ is the exact ground state (\ref{HG}). For $\chi<1$, the
difference increases with $\chi$ reaching its maximum at $\chi=1$.
At $\chi>1$ it drops to zero,
i.e. the narrow minimum "light" phase is resistant against $\eta$.

Fig. 4. Displacement of phonons $2 $ in the "heavy" region
$\gamma_2^2$; it evidently  causes  the "cave" due to
reflection $\eta$ of the ground state energy in the "heavy" region (Fig.
3)

Fig. 5. (a) The gap in the plane of $\mu$ and $\chi$ for
$\Omega= 0.5$. While the critical coupling $\mu_H (\Omega)$
for I-M transition in the
"heavy" phase is very slightly dependent of $\chi$, $\mu_L (\Omega, \chi)$
in the "light" phase is $\chi$ dependent.
(b) The shift of the metal-insulator transition in the "heavy" region
due to quantum fluctuations at strong coupling ($\mu> \mu_{H}$).

Fig. 6. The gap in the "light" region as function of $\Omega $ and $\mu$
for $\chi=1.5$. The fluctuations increase with $\Omega$.

Fig. 7. The competition of the level splitting and band width $T$ terms
($\tilde T= T M $ (\ref{havt})).
$\Delta_0$ is splitting of the levels (\ref{spl}), $\Delta_1$ neglects
the off-diagonal terms $H_{T_2}, H_{T_4}$,
(\ref{D1}).  $\Delta $ is the exact gap given by (\ref{gap}).

Fig. 8. Comparison of the phase diagram of the exact model (\ref{gap})
(solid line) with that one neglecting off-diagonal transfer (dashed line).
The metallic phase occurs on the left and the insulator on the right of the
respective line.

Fig. 9. Quantum fluctuations of (a) the  coordinate $\Delta^2 Q_1$,
(b) momentum $\Delta^2 P_1$ and (c) their product $\Delta ^2Q_1\Delta^2 P_1$.
(d) Detail of Fig. 3 underlying the region of parameters of Figs.
(a)-(c).

\end{document}